# Boolean and Non-Boolean Computation With Spin Devices

Mrigank Sharad, Charles Augustine, Georgios Panagopoulos, Kaushik Roy
Department of Electrical and Computer Engineering, Purdue University, West Lafayette, IN, USA
{msharad, caugust, gpanagop, kaushik}@.purdue.edu

**Abstract:** Recently several device and circuit design techniques have been explored for applying nano-magnets and spin torque devices like spin valves and domain wall magnets in computational hardware. However, most of them have been focused on digital logic, and, their benefits over robust and high performance CMOS remains debatable. Ultra-low voltage, current-mode operation of magneto-metallic spin torque devices can potentially be more suitable for non-Boolean logic like neuromorphic computation, which involve analog processing. Device circuit co-design for different classes of neuromorphic architectures using spin-torque based neuron models along with DWM or other memristive synapses show that the spin-based neuromorphic designs can achieve 15X-100X lower computation energy for applications like, image processing, data-conversion, cognitive-computing, associative memory and programmable-logic, as compared to state of art CMOS designs.

**Logic design with spin devices:** Until recently, nano-magnet logic (NML) was the only predominant spin-based computation scheme under exploration [1], [2]. It employs dipolar coupling between nano-magnets to perform logic computation and offers interesting features like non-volatility, zero leakage and compactness [1]. However, magnetic field based Bennett clocking used in NML requires pulsed current transmission through metal lines that makes it inefficient in terms of computation energy [2]. Theoretical possibility of alternate strategy for Bennett clocking in NML have been proposed recently [3], that makes use of anisotropic strain induced by multiferroic layers to turn magnets to hard axis. If successful, such a scheme could boost up the prospects of NML scheme and would make it attractive for low performance electronics like those used in biomedical implants.

Recent experiments on spin torque in device structures like lateral spin valve (LSV) [4], [5] (fig. 1a), domain wall magnets (DWM) [6], [7], and magnetic tunnel junctions, have opened new avenues for spin based computation. Several logic schemes have been proposed using such devices. Hybrid design schemes using MTJ have been explored that aim to club memory with logic and can possibly benefit from reduced memory-data traffic [12]. However, it might need thorough system level analysis to gauge the real benefits of such a strategy [13].

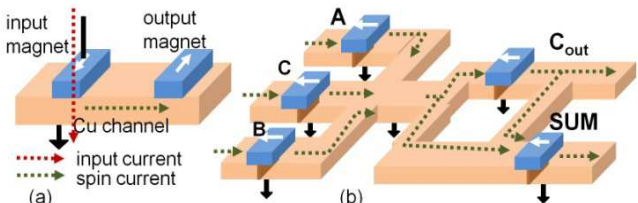

Fig. 1(a) Lateral spin valve with non-local spin injection, (b) ASL full adder based on spin majority evaluation.

Use of spin-torque in LSV's facilitated higher degree of spin current manipulation for logic. All spin logic (ASL) proposed in [8], employs cascaded LSV's interacting through spin torque, to realize logic gates and larger blocks like compact full adders [9], based on spin majority evaluation (fig. 1). It was suggested that maximization of spin mode processing of data and avoiding frequent conversion between spin to charge could lead to higher energy efficiency [8]. Analysis presented in [9] suggested the use of clocking in ASL circuits for lower computation energy (fig. 2). It was also shown that current-mode Bennett clocking in ASL could achieve speed-performance comparable to CMOS. The main bottle neck of this scheme stems from limited spin diffusion length of metal channels interconnecting the nano-magnets, which restricts the distance over which spin signal can be reliably transmitted [1]. This limits the fan-out of logic gates and makes the layout of arbitrary logic blocks challenging. Future prospects of ASL would strongly depend upon innovation and research in device materials that can improve the efficiency of spin-mode transport [14].

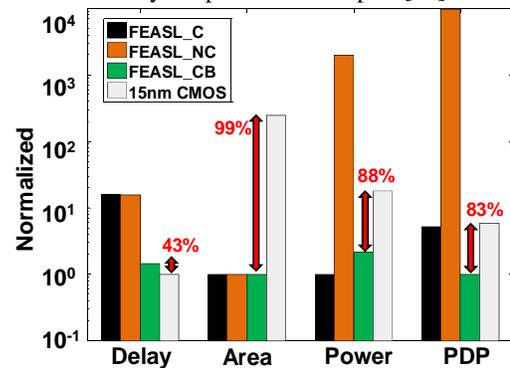

Fig. 2 Performance metrics for three different genres of ASL for a discrete cosine transform unit: clocked (FEASL_C), non-clocked (FEASL_NC), and ASL with current mode Bennett clocking (FEASL_CB). The results show ~83% lower computation energy for ASL_CB as compared to 15nm CMOS.

A number of logic schemes have been proposed based on current driven domain wall motion in magnetic nano-strips [10], [11]. Recently it has been shown that domain wall motion can be achieved with relatively small current density ($10^7$ A/cm$^2$) in magnetic nano-strips with perpendicular magnetic anisotropy [15]. This phenomenon was exploited in a recent proposal on DWM based logic scheme that employed short magnetic nano-wires to model logic gates [11]. Owing to non-volatility of DWM based logic cells fine-grained pipelining can be achieved in such a scheme leading to high throughput along with low computation energy for low and medium frequency data processing. However, multi-phase clocking employed in such schemes would necessitate the use of CMOS switches and hence the associated overhead needs to be considered in performance evaluation [14].

**Spin Devices for Neuromorphic computation - a Non-Boolean perspective:** Most of the spin based computation schemes proposed so far have been centered on modeling digital logic gates using these devices. A wider perspective on application of spin torque devices, however, would involve, not only exploring possible combination of spin and charge devices but, searching for computation models which can derive maximum benefits from such heterogeneous integration.

It can be noted that ultra low voltage, current-mode operation of magneto-metallic devices like LSV's and DWM's can be used to realize analog summation/integration and thresholding operations with the help of appropriate circuits, and, can be used to model energy efficient "neurons" [16]-[21]. Such device-circuit co-design can lead to ultra low power neuromorphic computation architectures, suitable for different data processing applications. The proposed hybrid design scheme can open a new frontier for spin torque based analog and digital computing.

**Neuron Model based on Spin Transfer Torque:** In the following subsection we present models for neurons based on lateral spin valves and domain wall magnets.

*A. Neuron Models Based on LSV*

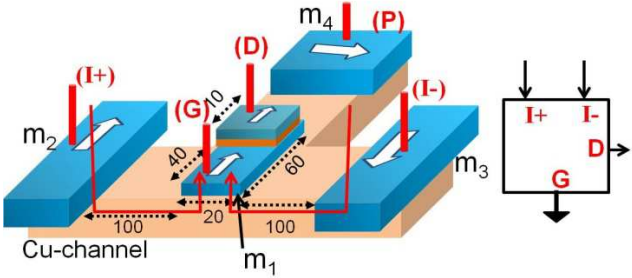

Fig. 3. Bipolar Spin Neuron (BSN) with local spin injection [4] (all dimensions in nm) : input terminals( I+/-) receive synapse currents, hard-axis biasing current is injected through the terminal *P*, terminal *D* is used to detect the state of the MTJ, *G* is the ground terminal through which current exits the device.

Transfer-function of a neuron can be expressed as the sign-function of weighted sum of inputs: $Y = 0.5(\text{sign}(\sum w_i \cdot I_i)+1)$, where the individual weights, $w_i$, can be either positive (excitatory) or negative (inhibitory) and inputs $I_i$ and the output $Y$ can acquire binary levels, one or zero. Fig. 3 shows the device structure for bipolar spin neuron [15], [18] that achieves this functionality. It constitutes of two opposite polarity input magnets $m_2$, $m_3$, connected to an output magnet $m_1$ through metal channel. All the excitatory and inhibitory inputs of the neuron connect to $m_2$ and $m_3$ respectively. At the beginning of the evaluation phase, a current pulse input through a preset magnet $m_4$, switches the output magnet, $m_1$, along its hard-axis [8]. The preset pulse is overlapped with the synchronous input current pulses received through the two anti-parallel input magnets. After removal of the preset pulse, $m_1$ switches back to its easy-axis. The final spin-polarity of $m_1$ depends upon the sign of the difference $\Delta I$, between the current inputs through $m_2$ and $m_3$. The lower limit on the magnitude of $\Delta I$ (hence, on current per-input for the neuron), for deterministic switching, is imposed by the thermal-noise in the output magnet, and, imprecision in Bennett-Clocking (BC). The effects of these non-idealities have been included in device simulation [18].

Some structural variations of LSV based neuron model have been discussed in [21]. For instance, a unipolar neuron with a single input magnet proposed in [21] might be more suitable for large networks, as the input to it is the small difference between positive and negative synapse currents rather than their sum. A compact spin-mode neuron-synapse unit using multi-input lateral spin valve (LSV) was also proposed, which employed programmable inputs in the form of DWM [16]. In [18] we showed that both local as well as non-local spin torque can be used to realize the neuron models based on LSV described above.

*B. Neuron Model Based on Domain Wall magnets*

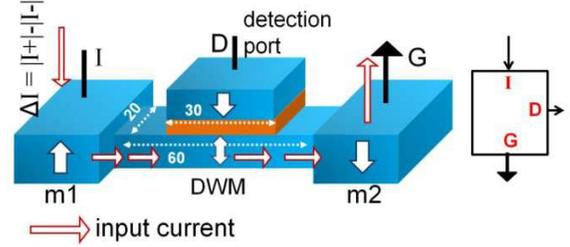

Fig.4 Unipolar spin neuron using domain wall magnet. :

Low current threshold for domain wall motion in Perpendicular Magnetic Anisotropy (PMA) nano-magnet strips [15], can be exploited to model a 'unipolar' neuron shown in fig 4 [21]. It constitutes of a thin and short (20x60x2 nm$^3$) DWM nano-strip connecting two antiparallel magnets of fixed polarity, $m_1$ and $m_2$. The magnet $m_1$ forms the input port, whereas, $m_2$ is grounded. Spin-polarity of the DWM layer can be written parallel to $m_1$ or $m_2$ by injecting a small current (~3µA) along it, depending upon the direction current flow [11]. MTJ based detection port is used for reading the spin polarity of the DWM stripe (fig. 4). Note that, application of such a structure in memory [15] and digital logic design [11] has been proposed earlier. We exploit this structure to model a neuron using appropriate circuit scheme. The input port of the DWM neuron receives the difference of the positive and the negative synapse currents, $\Delta I$. In addition to this, a bias current can be supplied which effectively shifts the DWM threshold closer to the origin. As a result, a small positive or negative $\Delta I$ (~1µA) can determine evaluation to one of the spin states, thereby realizing the sign function of a neuron. We employ dynamic CMOS latch for reading the MTJ, which results in only a small transient current drawn from the ground terminal (*G*) of the DWM neuron, which can be kept below its switching threshold. Additionally, the time domain threshold for domain wall motion also helps in preventing read disturb from the small transient current [10].

Note that the neuron models discussed above are applicable to clock synchronized neural networks, where all input signals of a neuron arrive concurrently during a clock phase. Step-wise motion of domain wall in longer nano-magnet stripes can be used to model 'integrating' neurons, applicable to asynchronous neuromorphic architectures, like spiking neural networks [21].

**Circuit Integration Scheme:** In this section, we describe the circuit integration scheme used in this work that exploits the ultra low voltage operation of the proposed spin neurons for energy efficient, analog-mode neuromorphic computation.

A dynamic CMOS latch senses the state of the neuron MTJ while injecting only a small transient current into the detection terminal [15]. The latch drives transistors operating in deep triode region, which transmit synapse current to all the fan-out neurons (fig. 5). The inter-connection scheme is different for unipolar and bipolar neuron models described in the previous section. For the bipolar neurons, two voltage levels differing by $\Delta V$ are used, i.e., *V* and $V+\Delta V$ (fig. 7a). Here *V* is a DC level close to 1V, whereas, $\Delta V$ can be around ~20mV. The source terminal of the output transistors are biased a $V+\Delta V$, where as the ground terminals of the receiving neurons are connected to *V*. Hence, the synapse currents, involved in computation, flow across a small terminal voltage Δ*V*, thereby, reducing the static power consumption resulting from large number of analog-mode synaptic communications in a neural network

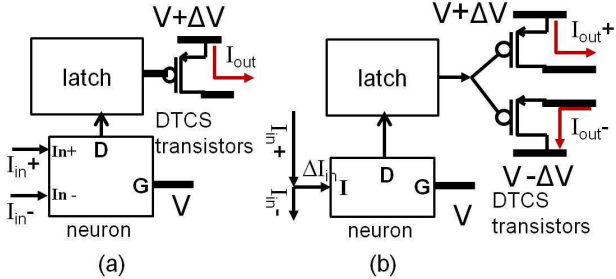

Fig. 5 Circuit integration scheme for (a) bipolar neurons and (b) unipolar neurons. (DTCS: deep triode current source transistos)

For the unipolar neurons, the currents received from negative and positive synapses need to be subtracted in charge mode, outside the device. This necessitates the use of three different voltage levels (fig. 5b). The transistors corresponding to positive weights, effectively source current to the receiving neurons ($I_{out}$+), whereas the transistors corresponding to the negative weights act as drains ($I_{out}$-). In this scheme, most of the current flows between the two extreme levels, *V*+Δ*V* and *V*- Δ*V*, whereas, only a small net current flows to and from the mid DC level *V*, through the neuron devices. Hence, routing the additional mid DC level may not be a significant design overhead. However, as the synapse currents in this case flow across 2Δ*V* , for a given strengths of the current source transistors, this scheme leads to 2X higher computation energy as compared to the case of bipolar neuron. Note that, we have chosen two relatively high DC levels differing by Δ*V* (/2Δ*V*), rather than small absolute levels +/-Δ*V* (+/-2Δ*V*), in order to ensure stable supply voltages [16].

**Design Example:** The circuit integration scheme described above can be employed for realizing different classes of neuromorphic architectures. Weights or connection strength between neurons can be realized in different ways. For the multi-input neuron proposed in [16], the DWM inputs act as compact spin-mode synapses. For other neuron models, weighted source transistors can be used for fixed, non-programmable designs [20]. Fig. 6 depicts a network of DWM neurons, based on this scheme and its analogy to a biological neural network.

Using a similar technique, we presented the design of cellular neural network (CNN) [20], that employs neighborhood connectivity and recursive operation (fig. 7). The weight of synaptic connectivity between neighbors and hence, the strength of the corresponding source transistors were set according to weight-templates for different image processing applications. Simulation results for some common image processing applications like edge extraction, motion detection, half-toning and digitization (fig. 8), using the spin based CNN, showed ~100x lower computation energy, as compared to state of art mixed-signal CMOS designs. As mentioned before, the main advantage comes from ultra low voltage, pulsed operation of spin neurons that are applied to analog computation.

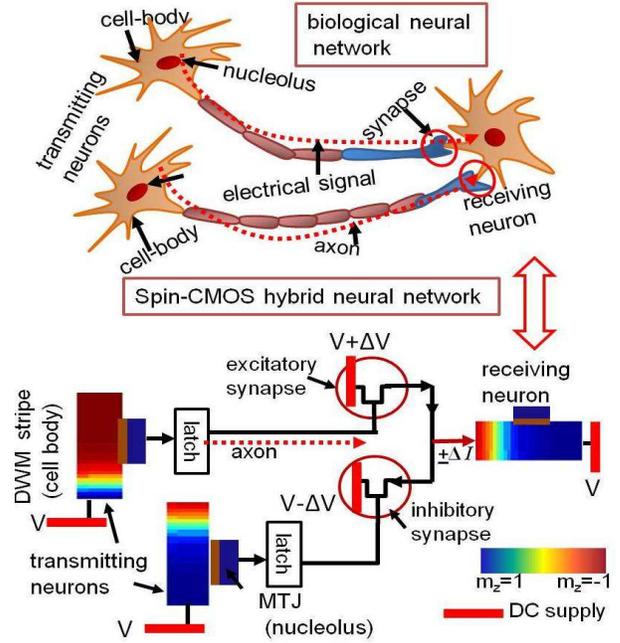

Fig. 6 Emulation of neural network using spin-CMOS hybrid circuit: In each neuron, the MTJ acts as the firing site, i.e., the nucleolus; DWM stripe can be compared to cell body and its spin polarization state is analogous to electrochemical potential in the neuron cell body which affects 'firing', the CMOS detection unit can be compared to axon that transmits electrical signal to the receiving neuron, and finally a weighted transistor acts as synapse as it determines the amount of current injected into a receiving neuron.

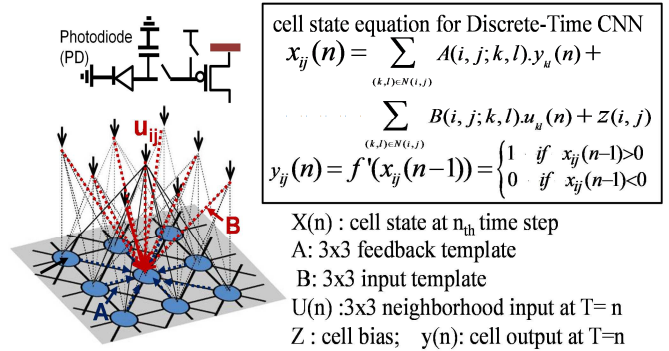

cell state equation for Discrete-Time CNN
$$x_{ij}(n) = \sum_{(k,l) \in N(i,j)} A(i,j;k,l).y_{kl}(n) + \sum_{(k,l) \in N(i,j)} B(i,j;k,l).u_{kl}(n) + z(i,j)$$
$$y_{ij}(n) = f'(x_{ij}(n-1)) = \begin{cases} 1 & if\ x_{ij}(n-1)>0 \\ 0 & if\ x_{ij}(n-1)<0 \end{cases}$$

X(n) : cell state at $n_{th}$ time step
A: 3x3 feedback template
B: 3x3 input template
U(n) :3x3 neighborhood input at T= n
Z : cell bias;  y(n): cell output at T=n

Fig. 7. 3x3 neighborhood architecture of CNN and equation for neuron's state: Current from each photosensor $u_{ij}$, is transmitted to 3x3 neighbors through transistors weighted according to *B* template, whereas, inter-neruron synaptic connectivity is realized with traistors weighted according to *A* template.

Programmable and self-adaptive weights can be realized using programmable conductive elements, like $TiO_2$ memristor and phase change memory (PCM) [17]. Fig. 9 shows a cross-bar neural network architecture using memristor (/PCM) synapses and bipolar spin neurons. Depending upon the polarity of the connectivity between an input line and a neuron, one of the two memristive junctions between them is driven to off state, while the other is programmed to match the required weight magnitude. The spin-neurons facilitate ultra-low voltage, pulsed synaptic communication across the cross-bar metal interconnects, thereby reducing the static-power consumption resulting from large number of inter-neuron signals per-cycle in a large-scale array. Such a design can provide ultra low power solution to several interesting applications, like, logic in memory, associative memory, programmable logic and pattern matching. Spiking neural networks based on memristive cross-bar arrays can realize self-learning networks for cognitive computing. Such a design employs some additional control circuits in each neuron to implement synaptic weight modification according to specific learning rules. But, most of the power consumption in all such networks results from synaptic communication, which can be reduced using DWM based integrating-neurons.

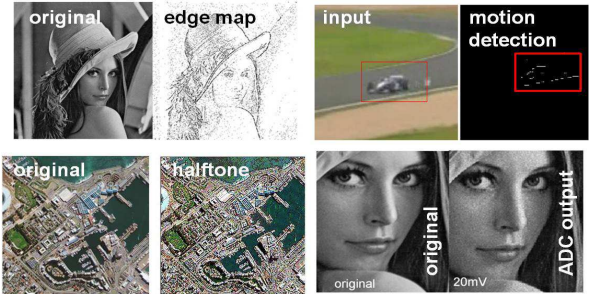

Fig. 8 CNN output for different image processing applications: feature extraction , motion detection , halftoning and digitization.

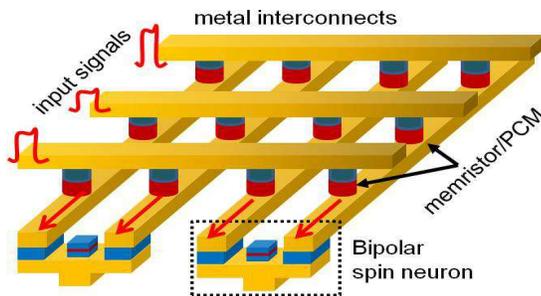

Fig. 9 Cross-bar network design using (a) unipolar spin nueron, (b) using bipolar spin neuron

**Simulation Framework :**

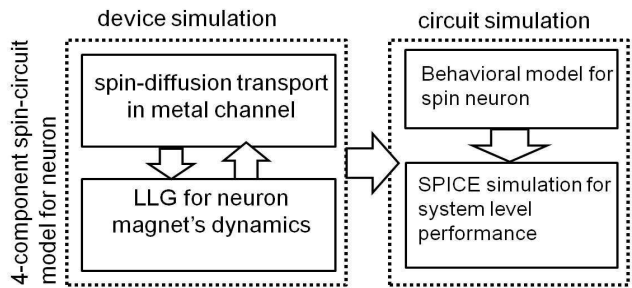

Fig. 8. Device circuit co-simulation framework employed in this work

Fig. 8 pictorially depicts the device-circuit co-simulation framework employed in this work to assess the system level performance for different neuromorphic architectures. The spin-diffusion transport was modelled using 4 component spin models of [22]. The device models for neurons have been benchmarked with experimental data on LSVs' [4,5] and DWM [7]. The corresponding behavioural models are used for circuit and system level simulation. Effect of parameter variation in spin neurons as well as CMOS has been incorporated in simulations.

Conclusion: Emerging spin device phenomena have lead to interesting prospects for low energy computation for Boolean as well as non-Boolean scheme. We noted that ultra low voltage, current mode operation of magneto-metallic spin torque-devices can be suitable for non-boolean, analog-mode computing. Using mixed mode simulation-framework, highly promising estimates were obtained for common data processing applications that show upto two orders of magnitude improvement in computation energy as compared to state of art CMOS design. The proposed hybrid design scheme can open a new frontier for spin torque based computing.

**Acknowledgement:** This research was funded in part by Nano Research Initiative and by the INDEX center. The authors thank S. Datta, B. Behin-Ain, and A. Sarkar for helpful discussions.